\documentclass{article}
\usepackage{longtable}
\usepackage{float}
\usepackage{subfloat}
\usepackage{lscape,epsfig}
\usepackage{graphicx}
\usepackage{amssymb}
\usepackage{amsmath}
\usepackage{natbib}
\textheight=22cm \DeclareSymbolFont{ppa}{OT1}{ppl}{m}{it}
\DeclareMathSymbol{\vv}{\mathalpha}{ppa}{'166}

minus 2.3mu \thickmuskip = 2.6mu plus 2mu minus 2.6mu

\let\svthefootnote\thefootnote

\begin{document}

\newcommand{\dd}{\,{\rm d}}
\newcommand{\ie}{{\it i.e.},\,}
\newcommand{\etal}{{\it $et$ $al$.\ }}
\newcommand{\eg}{{\it e.g.},\,}
\newcommand{\cf}{{\it cf.\ }}
\newcommand{\vs}{{\it vs.\ }}
\newcommand{\zdot}{\makebox[0pt][l]{.}}
\newcommand{\up}[1]{\ifmmode^{\rm #1}\else$^{\rm #1}$\fi}
\newcommand{\dn}[1]{\ifmmode_{\rm #1}\else$_{\rm #1}$\fi}
\newcommand{\upd}{\up{d}}
\newcommand{\uph}{\up{h}}
\newcommand{\upm}{\up{m}}
\newcommand{\ups}{\up{s}}
\newcommand{\arcd}{\ifmmode^{\circ}\else$^{\circ}$\fi}
\newcommand{\arcm}{\ifmmode{'}\else$'$\fi}
\newcommand{\arcs}{\ifmmode{''}\else$''$\fi}
\newcommand{\MS}{{\rm M}\ifmmode_{\odot}\else$_{\odot}$\fi}
\newcommand{\RS}{{\rm R}\ifmmode_{\odot}\else$_{\odot}$\fi}
\newcommand{\LS}{{\rm L}\ifmmode_{\odot}\else$_{\odot}$\fi}

\newcommand{\Abstract}[2]{{\footnotesize\begin{center}ABSTRACT\end{center}
\vspace{1mm}\par#1\par \noindent {~}{\it #2}}}

\newcommand{\TabCap}[2]{\begin{center}\parbox[t]{#1}{\begin{center}
  \small {\spaceskip 2pt plus 1pt minus 1pt T a b l e}
  \refstepcounter{table}\thetable \\[2mm]
  \footnotesize #2 \end{center}}\end{center}}

\newcommand{\TableSep}[2]{\begin{table}[p]\vspace{#1}
\TabCap{#2}\end{table}}

\newcommand{\FigCap}[1]{\footnotesize\par\noindent Fig.\  %
  \refstepcounter{figure}\thefigure. #1\par}

\newcommand{\TableFont}{\footnotesize}
\newcommand{\TableFontIt}{\ttit}
\newcommand{\SetTableFont}[1]{\renewcommand{\TableFont}{#1}}
\newcommand{\MakeTable}[4]{\begin{table}[htb]\TabCap{#2}{#3}
  \begin{center} \TableFont \begin{tabular}{#1} #4
  \end{tabular}\end{center}\end{table}}

\newcommand{\MakeTableSep}[4]{\begin{table}[p]\TabCap{#2}{#3}
  \begin{center} \TableFont \begin{tabular}{#1} #4
  \end{tabular}\end{center}\end{table}}

\newenvironment{references}%
{ \footnotesize \frenchspacing
\renewcommand{\thesection}{}
\renewcommand{\in}{{\rm in }}
\renewcommand{\AA}{Astron.\ Astrophys.}
\newcommand{\AAS}{Astron.~Astrophys.~Suppl.~Ser.}
\newcommand{\ApJ}{Astrophys.\ J.}
\newcommand{\ApJS}{Astrophys.\ J.~Suppl.~Ser.}
\newcommand{\ApJL}{Astrophys.\ J.~Letters}
\newcommand{\AJ}{Astron.\ J.}
\newcommand{\IBVS}{IBVS}
\newcommand{\PASP}{P.A.S.P.}
\newcommand{\Acta}{Acta Astron.}
\newcommand{\MNRAS}{MNRAS}
\renewcommand{\and}{{\rm and }}
\section{{\rm REFERENCES}}
\sloppy \hyphenpenalty10000
\begin{list}{}{\leftmargin1cm\listparindent-1cm
\itemindent\listparindent\parsep0pt\itemsep0pt}}%
{\end{list}\vspace{2mm}}

\def\TYLDA{~}
\newlength{\DW}
\settowidth{\DW}{0}
\newcommand{\dw}{\hspace{\DW}}

\newcommand{\refitem}[5]{\item[]{#1} #2%
\def\REFARG{#3}\ifx\REFARG\TYLDA\else, {\it#3}\fi
\def\REFARG{#4}\ifx\REFARG\TYLDA\else, {\bf#4}\fi
\def\REFARG{#5}\ifx\REFARG\TYLDA\else, {#5}\fi.}

\newcommand{\Section}[1]{\section{#1}}
\newcommand{\Subsection}[1]{\subsection{#1}}
\newcommand{\Acknow}[1]{\par\vspace{5mm}{\bf Acknowledgements.} #1}
\pagestyle{myheadings}

\newfont{\bb}{ptmbi8t at 12pt}
\newcommand{\xrule}{\rule{0pt}{2.5ex}}
\newcommand{\xxrule}{\rule[-1.8ex]{0pt}{4.5ex}}

\begin{center}
{\Large\bf
 The Clusters AgeS Experiment (CASE).  \\
 Variable stars in the field of  
 the globular cluster NGC 3201}{\LARGE$^\ast$}
 \vskip1cm
  {\large
      ~~J.~~K~a~l~u~z~n~y$^1$\dag
      ~~M.~~R~o~z~y~c~z~k~a$^1$,
      ~~I.~B.~~T~h~o~m~p~s~o~n$^2$,
      ~~W.~~N~a~r~l~o~c~h$^1$
      ~~B.~~M~a~z~u~r$^1$
      ~~W.~~P~y~c~h$^1$
      ~~and~~A.~~S~c~h~w~a~r~z~e~n~b~e~r~g~~-~~C~z~e~r~n~y$^1$
   }
  \vskip3mm
{ $^1$Nicolaus Copernicus Astronomical Center, ul. Bartycka 18, 00-716 Warsaw, Poland\\
     e-mail: (mnr, wnarloch, batka, pych, alex)@camk.edu.pl\\
  $^2$The Observatories of the Carnegie Institution for Science, 813 Santa Barbara
      Street, Pasadena, CA 91101, USA\\
     e-mail: ian@obs.carnegiescience.edu}
\end{center}

\vspace*{7pt}
\Abstract 
{The field of the globular cluster NGC 3201 was monitored between 1998 and 2009 in a search 
for variable stars. $BV$ light curves were obtained for 152 periodic or likely periodic 
variables, 57 of which are new detections. Thirty-seven newly detected variables are proper
motion members of the cluster. Among them we found seven detached or semi-detached eclipsing 
binaries, four contact binaries, and eight SX~Phe pulsators. Four of the eclipsing binaries 
are located in the turnoff region, one on the lower main sequence and the remaining two 
slightly above the subgiant branch. Two contact systems are blue stragglers, 
and another two reside in the turnoff region. In the blue straggler region a total of 266 objects 
were found, of which 140 are proper motion (PM) members of NGC 3201, and another 19 are field stars. 
Seventy-eight of the remaining objects for which we do not have PM data are located within the 
half-light radius from the center of the cluster, and most of them are likely genuine blue 
stragglers. Four variable objects in our field of view were found to coincide with X-ray sources: 
three chromosperically active stars and a quasar at a redshift $z\approx0.5$. 
}
{globular clusters: individual (NGC 3201) -- stars: variables -- 
stars: SX Phe -- blue stragglers -- binaries: eclipsing
}

\let\thefootnote\relax\footnotetext{\dag Deceased}
\let\thefootnote\relax\footnotetext
{$^{\mathrm{\ast}}$Based on data obtained with the Swope telescope at Las Campanas Observatory.}
\let\thefootnote\svthefootnote

\Section{Introduction} 
\label{sec:intro}
NGC 3201 is a nearby ($d=4.9$ kpc) globular cluster (GC), projected against 
the edge of the galactic disk at $l=277^\circ.2$ and $b=8^\circ.6)$ in a field
with an average reddening of $E(B-V)=0.24$ mag. Its core radius, half-light radius, 
concentration parameter, [Fe/H] index, and radial velocity are equal to 1$'$.3, 
3$'$.1, 1.29, -1.59 and +494.4$\pm$0.2 km/s, respectively (Harris 1996, 2010 edition). 
The exceptional kinematic properties
of NGC 3201 (extreme radial velocity and retrograde orbit around the center of Milky 
Way) strongly suggest an extragalactic origin for this object. However, as it exhibits 
no peculiarities in the chemical composition, its chemical evolution must have been 
similar to that of other galactic GCs (Mu\~noz et al. 2013). The proximity and low 
concentration make NGC 3201 an attractive target for detailed studies with ground-based 
telescopes. Unfortunately, photometric observations are seriously hampered by large 
differential reddening, with $E(V-I)$ varying by up to 0.2~mag on a scale of arcminutes 
(von Braun \& Mateo 2000). 

Early pre-CCD searches for variables in the field of NGC 3201, summarized by Clement 
et al. (2001), Mazur et al. (2003, hereafter M03), and Arellano Ferro et al. (2014), 
resulted in the detection of 100 objects, 
some of which were found to be nonvariable in later studies. The four CCD surveys that have been
performed so far (von Braun \& Mateo 2002, M03, Layden \& 
Sarajedini 2003, and Arellano Ferro et al. 2014) reported an additional 24 discoveries. Altogether, 
113 variable stars have been listed,  consisting of 86 RR Lyr and 16 SX Phe pulsators, 3 eclipsing 
binaries, and 8 long-period/irregular variables). All of these presumed to be cluster 
members.  

The 57 new variables presented in this contribution are a result of the long-term photometric 
survey conducted within the CASE project (Kaluzny et al. 2005) using telescopes of the 
Las Campanas Observatory. Section~2 contains a brief report on the observations and explains the 
methods used to calibrate the photometry. Newly discovered variables are presented 
and discussed in Section~3, and the paper is summarized in Section~4.
\section{Observations}
\label{sec:obs}
Our paper is based on images acquired with the 1.0-m Swope telescope using the 
$2048\times 3150$ SITE3 camera. The field of view was $14.8\times 22.8$ arcmin$^2$ 
at a scale of 0.435 arcsec/pixel. Observations were conducted on 39 nights from 
January 5, 1998 to March 25, 2009. The same set of filters was always used. For the 
analysis 975 $V$-band images and 112 $B$-band images were selected. The seeing ranged from 
1$''$.2 to 3$''$.6 and 1$''$.2 to 2$''$.1 for $V$ and $B$, respectively, with median 
values of 1$''$.5 and 1$''$.4. 

\begin{figure}[H]
   \centerline{\includegraphics[width=0.95\textwidth,
               bb = 27 288 560 560, clip]{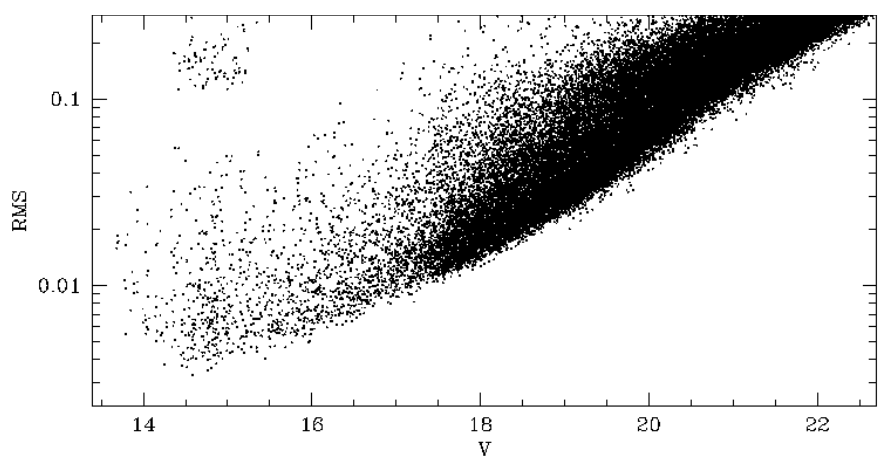}}
   \caption{ Standard deviation vs. average $V$ magnitude for
    light curves of stars from the NGC 3201 field. The points spread
    around $RMS$ $\approx$ 0.15 and $V$ $\approx$ 15 mag are RR Lyr stars. 
    \label{fig:rms}}
\end{figure}

The photometry was performed using an image subtraction technique implemented in the 
DIAPL package.\footnote{Available from http://users.camk.edu.pl/pych/DIAPL/index.html} 
To reduce the effects of PSF variability, each frame was divided into 6$\times$4  
overlapping subframes. The reference frames were constructed by combining seven images 
in $V$ and five in $B$ with an average seeing of 1.$''$18 and 1.$''$20, respectively.
The light curves derived with DIAPL were converted from differential counts to magnitudes 
based on profile photometry and aperture corrections determined separately for each 
subframe of the reference frames. To extract the 
profile photometry from reference images and to derive aperture corrections, the 
standard Daophot, Allstar and Daogrow (Stetson 1987, 1990) packages were used. The 
profile photometry was also extracted for each individual image, enabling useful 
photometric measurements of stars which were overexposed on the reference frames, 
and facilitating the identification of variable objects in crowded fields (which is 
sometimes problematic when image subtraction alone is used). As shown in Fig. 
\ref{fig:rms}, the accuracy of our photometry decreases from $\sim$4 mmag at $V=14.5$~mag 
to 40 mmag at $V=19.5$ mag and 100 mmag at $V=20.5$ mag.
\subsection{Calibration}

\begin{figure}[H]
   \centerline{\includegraphics[width=0.95\textwidth,
               bb = 25 536 562 687, clip]{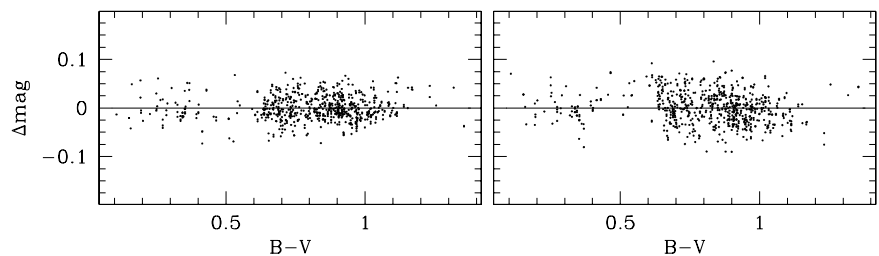}}
   \caption{Differences between our photometry transformed to the standard system 
    and that of M03. $V$-magnitudes and $B-V$ colors of 679 
    stars are compared in left and right panel, respectively. 
    \label{fig:calib}}
\end{figure}

\begin{figure}
   \centerline{\includegraphics[width=0.95\textwidth,
               bb = 56 200 560 560, clip]{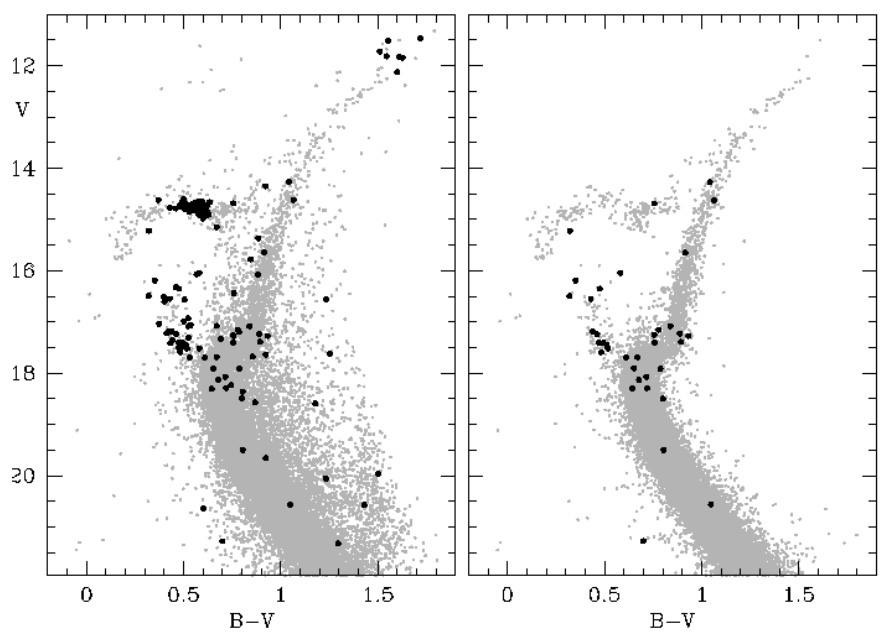}}
   \caption{CMD of NGC 3201. Left: all stars within our field of view for which 
    proper motions were measured. Black points mark all variables detected within 
    the present survey. Right: PM-members of the cluster only. Black points mark the 
    newly detected variable members of NGC 3201.
    \label{fig:cmds}}
\end{figure}

M03 performed a regular transformation to the $UBV$ system using their observations 
of Landolt standards (Landolt 1992). Since we did not have such data collected with 
the SITE3 camera, we decided to use their stars as secondary standards. Specifically, 
679 stars with formal errors in $V$ smaller than 0.015 mag were chosen, spanning a range 
14 -- 18 mag in $V$ and 0.15 - 1.35 mag in $B-V$. 
The linear transformations 
\begin{align}
V &= 2.6621(3) -0.0013(9)*(b-v)\\
B-V&= 0.5521(3) +1.0262(9)*(b-v)
\end{align}
proved to be entirely adequate, second-order 
terms did not improve the fit in any significant way. Residuals from the fit, defined 
as our transformed value minus the corresponding M03 value, are shown in Fig. \ref{fig:calib}. 
Mean and RMS values of $\Delta V$ are 0.0002 and 0.025 mag, respectively, and 
no systematic dependence on the color is seen. Although mean and RMS values of $\Delta (B-V)$ 
are also small (0.001 and 0.025 mag, respectively), for $B-V > 0.9$ mag our colors are 
slightly bluer than those of M03. However, since the main goal of the present survey 
is to detect new variables, inaccuracies of this order are unimportant.
Fig.~\ref{fig:cmds}, based on the reference images, shows the color-magnitude 
diagram (CMD) of the observed field. To make the figure readable, only stars with measured proper 
motions (Narloch et al., in preparation) are  selected to serve as a background for the 
variables. Stars tagged as proper-motion (PM) members of the cluster are shown in the right frame.
\subsection{Search for variables}
The search for variable stars was conducted using the AOV and AOVTRANS
algorithms implemented in the TATRY code (Schwarzenberg-Czerny
1996, Schwar\-zenberg-Czerny \& Beaulieu 2006). We examined 
time-series photometric data of 43512 stars with $V<21.5$ mag. The limits of detectable 
variability depended on the accuracy of photometric measurements: the smallest 
amplitude we were able to measure was 0.016 mag for $14.5<V<15.5$ mag, and 0.35 mag 
for $18.5<V<20.5$ mag. We obtained light curves of all previously known variables within 
our field of view\footnote{Available from the CASE archive at http://case.camk.edu.pl},
and discovered 57 new variable or likely variable stars, 36 of which are PM-members of 
NGC~3201. Finding charts for the newly detected variables are shown in Figs.~\ref{fig:maps1} 
and \ref{fig:maps2}. For completeness, we also include the finding chart for the eclipsing 
binary V119 of Clement et al. (2001, 2012 
edition\footnote{http://www.astro.utoronto.ca/~cclement/cat/C1015m461}), whose light 
curve we decided to publish because we detected the previously unobserved secondary minimum 
(as a result, the period of V119 turned out to be half or the previously reported value).

\section{The new variables}

Altogether, we discovered 49 periodic variables and found 8 stars suspected of variability. 
The basic data of this sample plus the above mentioned binary V119 are listed in Table \ref{tab:vardata}. 
We follow the naming convention of Clement et al. (2001) and Arellano Ferro et al. (2014). The 
list begins with V119, and, as the last previously cataloged variable was V124, it continues 
from V125 on. Stars 
V125 -- V160 are PM-members of NGC 3201, whereas VN1 -- VN21 are either nonmembers or their 
proper motions have not been determined. The equatorial coordinates in Table \ref{tab:vardata} 
conform to the UCAC4 system (Zacharias et al. 2013), and are accurate to about 0$''$.2. 

\begin{figure}
   \centerline{\includegraphics[width=0.95\textwidth,
               bb = 28 50 560 736, clip]{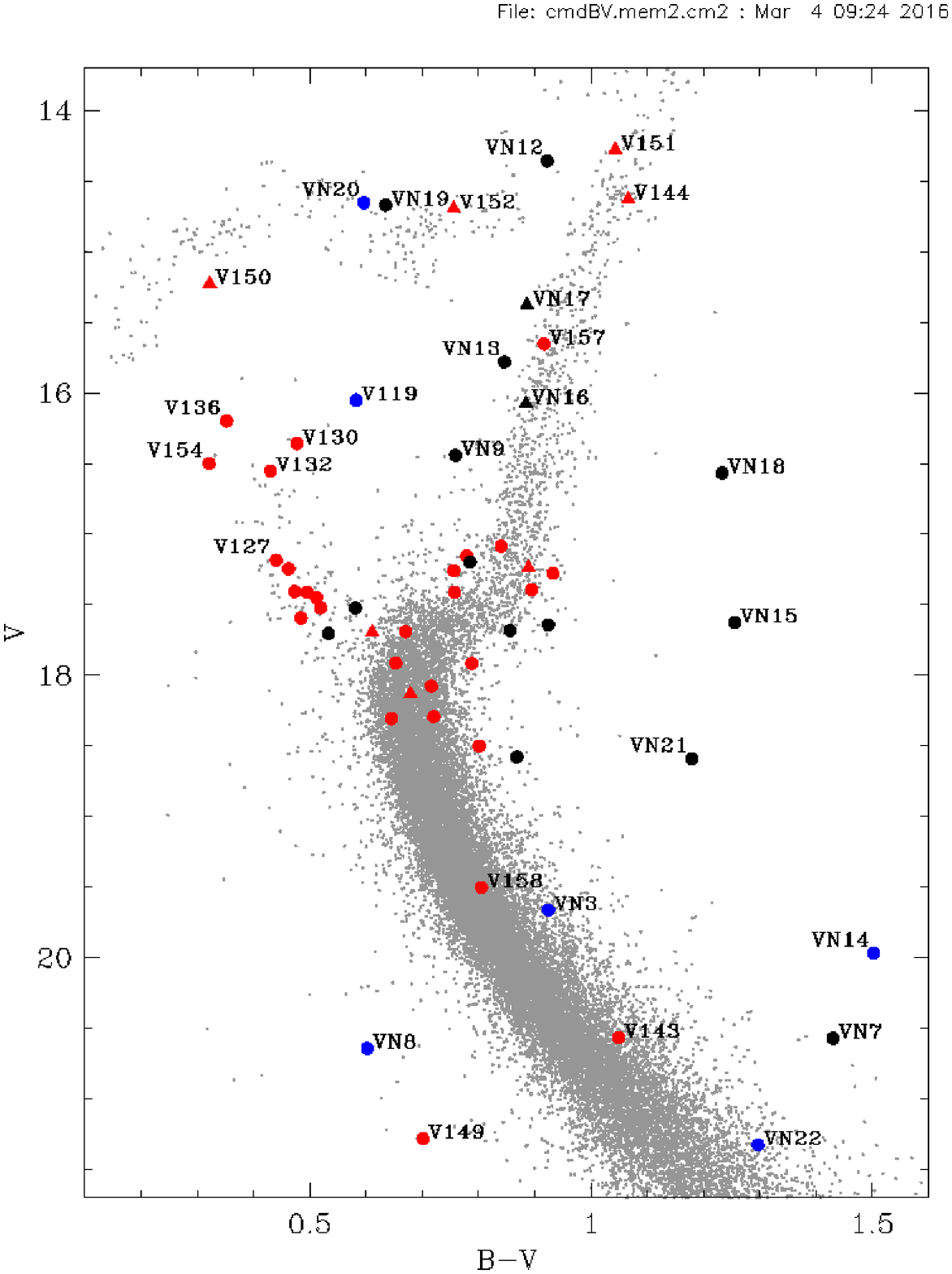}}
   \caption{Color-magnitude diagram for NGC 3201 with indicated locations of newly detected 
    variables. Red, black and blue symbols denote, respectively, members, nonmembers and 
    objects for which the PM-membership data is missing or ambiguous.
    Circles: periodic variables. Triangles: suspected variables.
    The gray background stars are the same as in the right frame of Fig.~\ref{fig:cmds}.
    The central part of the diagram is shown magnified in Fig.~\ref{fig:cmd_var2}.
    \label{fig:cmd_var1}}
\end{figure}

\begin{figure}
   \centerline{\includegraphics[width=0.95\textwidth,
               bb = 30 170 562 688, clip]{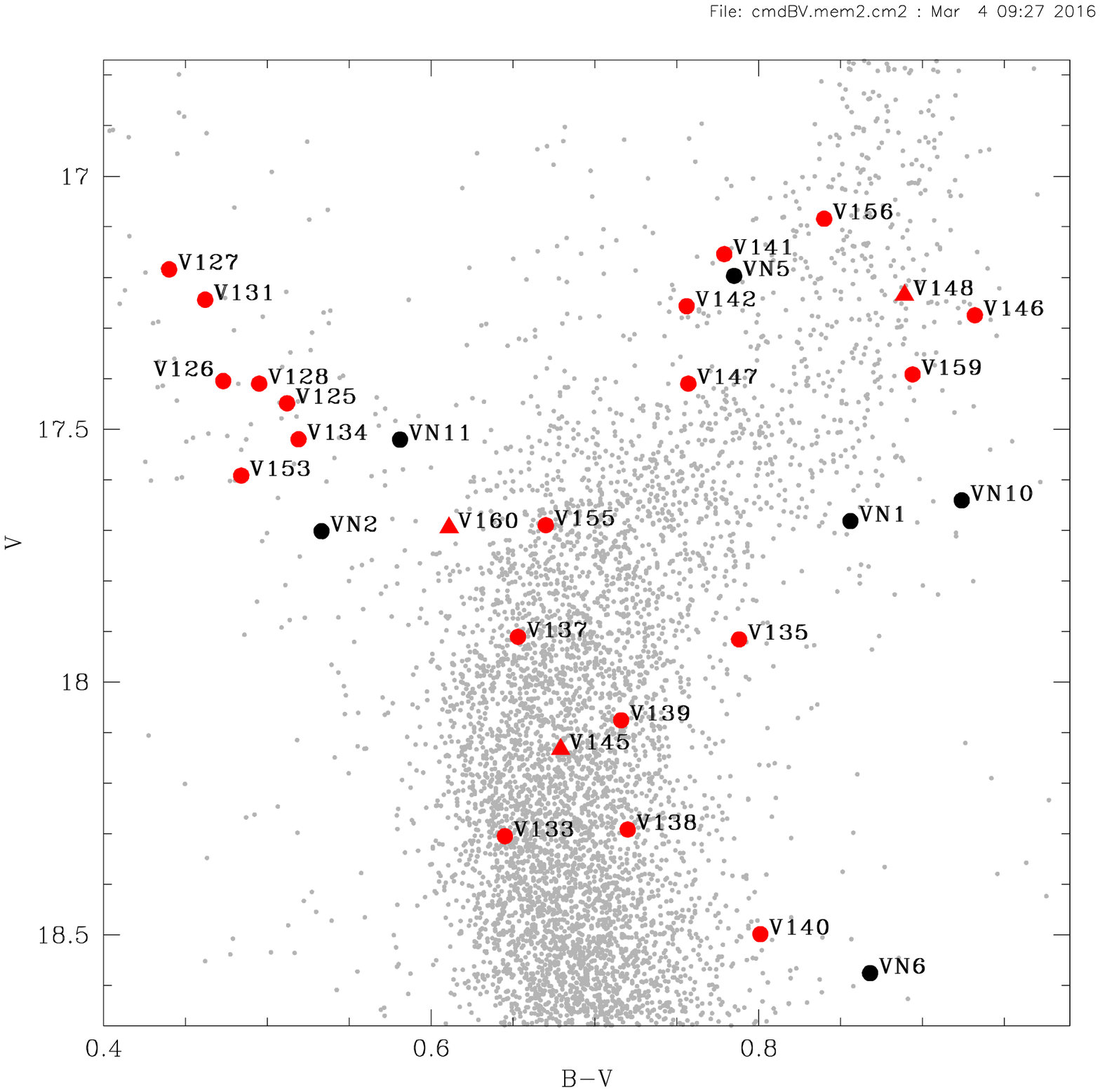}}
   \caption{The turnoff region from Fig. \ref{fig:cmd_var1}.  
    \label{fig:cmd_var2}}
\end{figure}

The $V$-magnitudes listed in Table \ref{tab:vardata} correspond to the maximum light in the case of eclipsing 
binaries, while for the remaining variables average magnitudes are listed. For each variable 
the $B-V$ color is given, followed by the amplitude in the $V$-band. Periods of variability were 
found for all stars except the suspected secular variable V160, and the eclipsing binaries V140 
and VN10 for which only a part of a single eclipse was observed. The last column of Table 1 gives 
the membership status based on proper motions from Zloczewski et al. (2012) and Narloch et al. 
(in preparation). Phased light curves of the variables from Table 1 are presented in 
Figs.~\ref{fig:curves1} -- \ref{fig:curves5}. The light curve of the W UMa system V133 could not 
be phased with a single period. This star is analyzed in the Appendix along with W UMa systems 2 
and 4 from the list of von Braun \& Mateo (2002) which show the same behavior (these latter two stars 
do not belong to NGC 3201 according to our PM data). Some light curves are clearly periodic in one 
or two seasons only. In these cases we give periods obtained for the season indicated in the 
footnote to Table 1. A CMD of the cluster with the locations of the variables is shown in 
Figs.~\ref{fig:cmd_var1} and \ref{fig:cmd_var2}. The gray background stars in the CMD are the 
PM-members of NGC 3201 from the right frame of Fig. \ref{fig:cmds}. Variables that are PM-members 
of the cluster are labeled in red, those with PM dataindicating that they are field objects in black, 
and those for which the PM data are missing or ambiguous in blue. 

V146, V147 and VN18 coincide with the X-ray sources X101736.16-462539.8, X101740.99-462142.1 and 3XMM 
J101806.9-462055 from the HEASARC Master Catalog (the corresponding coordinate differences are 0.04, 
0.50 and 1.17 arcsec, respectively). 
Webb et al. (2006), who observed NGC 3201 with the XMM-Newton telescope, report eight X-ray sources 
in the field of the cluster. Of these, source \#23 (= 3XMM~J101715.5-462253) coincides with object 
\#337214 on our list of light curves, located at $(\alpha, \delta)$ = (154.31503, -46.38146). 
Between May 2001 and June 2005 \#337214 brightened in $V$ by ~1.7 mag. We took its spectrum, and 
it turned out it was a quasar with $V_{max}\approx20.5$ mag, $B-V\approx0.5$ and $z\approx0.5$. 
As an extragalactic object, it is excluded from Table~\ref{tab:vardata} and Fig. \ref{fig:cmd_var1}.  

\subsection{Detached eclipsing binaries}
\label{subs:EA}

We detected 13 detached eclipsing binaries, of which eight are proper motion members of the 
cluster. Systems V138 -- V142, located in turnoff or subgiant region of the CMD, are interesting 
targets for detailed follow-up studies aimed at the determination of their absolute parameters
together with age and distance of NGC 3201. We are presently conducting such an analysis for V138, 
V139, V141 and V142. Their systemic velocities of 494.5, 498.3, 494.4, and 492.9 km s$^{-1}$, 
respectively, clearly confirm the membership status from Table \ref{tab:vardata}. At orbital periods of 9.20, 
8.79, 10.0 and 27.7 d, all four systems are well detached, so that their components should have 
evolved as single stars (provided, of course, that have not suffered from close encounter effects). 
Projected distances from the center of the cluster range from 0.2$r_h$ 
(V141) to 1.6$r_h$ (V142); $r_h$ being the half-light radius of 3$'$.1 (Harris 1996, 2010 edition). 
The light curves of V138, V139 and V142 are stable; that of V141 shows clear signs of a vigorous 
activity of at least one component. Orbits of V138 -- V141 are circular; that of V142 is strongly 
eccentric with $e=0.47$. As in this case the circularization time is longer than the Hubble time 
(e.g. Mazeh 2008; Mathieu et al. 2004), it is possible that the shape of the orbit has been 
preserved since the formation of NGC 3201 12~Gyr ago (Dotter et al. 2010). 
Accounting for photometry errors, the CMD-location of V140 is compatible with that of a system 
composed of stars with nearly the same mass. Since a part of one eclipse was only recorded, the 
orbital period is not known. If it is long enough, V140 will be the fifth system suitable for a 
detailed analysis of component parameters. 

The light curve of the turnoff binary V137 shows only one eclipse $\sim$0.1 mag deep. Along with 
its location close to the red edge of of the main sequence this suggests a rather large mass ratio 
of the components. The primary of this system is about to exhaust hydrogen in the core and enter the rapid 
expansion phase in which the mass from its envelope will be transferred to the smaller and redder 
secondary. Eventually, the secondary is likely to end up as another blue straggler. The period of 
V137 could not have been determined based on the data presented here - we took from the unpublished 
PhD thesis of BM. 
V143 is a tight lower main sequence binary
with a pronounced ellipsoidal effect, too faint for spectroscopic observations given the period 
of only 0.38 d. V144 exhibits low amplitude ($\Delta V = 0.02$ mag) variations with $P = 0.94$ d
that can be interpreted as arising from an eclipsing binary with a strong ellipsoidal effect.
If this is the case, then to explain its location near the tip of the Red Giant Branch (RGB) 
one has to assume that the 
binary is blended with a red giant. Thus, despite the relatively good photometric accuracy 
($\sim0.005$~mag, see Fig. \ref{fig:rms}), an additional confirmation of the variability of V144 
is needed.  
    
V119 is a likely PM-member of NGC 3201, whose projected location at the edge of the cluster's core 
additionally suggests the reality of its membership. This Algol-type system was discovered by 
Layden \& Sarajedini (2003, their star \#564), who however did not register the shallow secondary 
minimum, and deduced a period two times too long. Our nearly complete lightcurve (Fig. \ref{fig:curves1}) 
suggests a semidetached or nearly semidetached system very similar to the blue straggler V60 in M55, 
described in detail by Rozyczka et al. (2013). If radial velocity measurements confirm the membership
of V119, it will be the second example of a blue straggler ``in the making'', i.e. transferring 
hydrogen-rich matter from the less massive, but more evolutionary advanced secondary to the primary. 

VN8, for which we do not have PM data, is located within $r_h$. This blue system with a period 
of nearly three days and deep ($\sim0.8$ mag) eclipses must be composed 
of two similar subdwarfs. Seen nearly edge-on, it provides an excellent opportunity to obtain accurate 
parameters of its components. If radial velocity measurements confirm its membership, it will bring 
valuable information about late evolutionary stages of cluster binaries. 

VN6, VN7 and VN10 are located on the CMD so far redwards of the main sequence that they must be 
foreground or background objects. Also our PM data indicate that they do not belong to NGC 3201. 
The light curve of VN9 is characteristic of a vigorous chromospheric activity. On the CMD this system 
is placed slightly bluewards of the RGB, and its projected distance of 0.9$r_h$ from the center of the 
cluster suggests it might be a member of NGC 3201 despite our PM-measurement which seems to exclude 
this possibility.

\subsection {Contact binaries}
\label{subs:EW}

We identified nine variables with W UMa-type light curves of which four are PM-members of 
the cluster, and another three are field interlopers (for the remaining two systems no 
PM data are available). Small amplitudes and nearly sinusoidal variations of 
PM-members V134 -- V136 suggest that they may be ellipsoidal variables rather than genuine 
contact binaries. Spectroscopic data are needed to resolve this ambiguity. The PM-member 
V133 has a classical W UMa lightcurve with an amplitude of $\sim$0.4~mag and two total 
eclipses of  slightly different depths. Our attempts to phase the data with a single period 
failed, and an $O-C$ analysis showed that the period of this system is shortening (see 
Appendix for details). 

These four systems all reside within $r_h$, strongly suggesting cluster membership. None of 
them is located below the main sequence turnoff, providing another example in support of 
the general paucity of contact binaries on unevolved main sequences of globular clusters 
(Yan \& Mateo 1994; Kaluzny et al. 2014). At least in globular clusters, the principal 
factor enabling the formation of such systems from detached binaries seems to be 
nuclear evolution: a contact configuration is achieved once the more massive component 
reaches the turnoff and starts to expand, whereas the frequently invoked magnetic braking 
(e.g. Stepien \& Gazeas 2012, and references therein) plays only a minor role. A possible 
exception to this rule might be VN3, located slightly redwards of the lower main sequence and 
about 1.5 mag below the turnoff. Its light curve is asymmetric, with maxima of different 
heights and minima of different depths. This suggests that VN3 is a near-contact binary 
of the V1010 Oph subclass (Gu et al. 2004; Shaw 1994). Unfortunately, we do not have PM data for 
this object, and spectroscopy remains the only means to determine its membership status. 

Proper motions of VN1 and VN5 indicate that these systems do not belong to NGC 3201. The 
same conclusion is suggested by their position on the CMD, where they are flanking the base 
of the red giant branch (RGB). Most probably, they are foreground objects. 
VN2, a binary with a classical W UMa lightcurve, resides in the blue straggler region at a
projected distance of only 0.2 $r_h$ from the center of the cluster. Thus, despite the 
contradicting PM-evidence, we cannot exclude it being a member. Again, spectroscopy must 
decide. For the low-magnitude system VN4 $(\sim21\le V \le \sim22)$ we have $V$-photometry 
only, and we cannot locate it on the CMD. Its proper motion was also impossible to measure.  

\subsection{Variable stars among blue stragglers}
In the CMD of NGC 3201 there are 266 candidate blue stragglers (BSs) with $16.0<V<17.7$ and 
$0.23<B-V<0.60$. Of these 140 are PM-members or likely PM-members of NGC 3201, and 19 are field 
stars. For the remaining 107 BS-candidates no PM data are available, but 78 of them reside at a 
projected distance of less than $r_h$ from the center of the cluster. Assuming that the six 
candidates residing between 5$r_h$ and 7$r_h$ are in fact field interlopers, we may 
expect that 74 out of 78 candidates at $r<r_h$ are genuine BSs. Among the BS population, except 
the eclipsing binaries V119, V136, V136 and VN2 described in Subsections \ref{subs:EA} and 
\ref{subs:EW}, we found 26 periodic variables, ten of which are new detections. 
V153 and V154 exhibit more or less sinusoidal luminosity variations whose nature is difficult 
to establish based on available data. V125 -- V132 are SX Phe-type pulsators with periods 
ranging from 0.033 to 0.052 d.
Since V129 is located about half-arcsecond away from a blue star we were not able to measure 
its $B$-magnitude; however its $V$-magnitude, light curve and proper motion all unambiguously
determine its nature as a BS belonging to NGC3201.
Altogether, 24 SX Phe stars are now known in NGC 3201; many of them bimodal.
\subsection{Remaining objects}

The turnoff star V145 shows clear sinusoidal variations with $P=24.3$ d and an amplitude $\Delta 
V\sim0.1$ mag. However, since it is a blend (see Fig. \ref{fig:maps1}), we marked it as a suspected 
variable. V146, located at the RGB and exhibiting semi-regular variations with $\Delta V\sim0.15$~mag 
and $P=13.76$ d, is a spotted variable. Similar, but even 
less regular activity is observed in V147, another RGB object with a characteristic time scale of $\sim2$ d. 
This object, like V146, coincides with an X-ray source. For V148 a clear periodicity 
with $P=0.47$ d and $\Delta V\sim0.025$~mag was recorded only in our best observing season 
(May 2003). As such behavior is not expected for an RGB object, we classified it as a suspected 
variable. One of the most interesting new variables is V149, which in all seasons shows 
regular large-amplitude ($\Delta V\sim0.5$ mag) sinusoidal variations with a very well defined 
period of only 0.062 d. A period two times longer is also acceptable, but in either case the object is 
apparently a tight binary composed of compact stars, probably subdwarfs. Unfortunately, the short period 
makes spectroscopy of V149 a very challenging task. V150 -- V152 are bright stars with sinusoidal 
light curves of marginally detectable amplitudes and periods from 0.19 to 1.3~d. Since it is 
hard to give physical explanations of such behavior at their CMD-locations (horizontal branch and RGB
tip), we marked all of them as suspected variables. 

The turnoff object V155 with $P=0.25$ d and $\Delta V\sim0.05$~mag may be a W~UMa system seen at 
a low inclination angle. V156 is located at the RGB base, has $\Delta V\sim0.03$ mag, and a very 
stable period of only 0.041 d. Most probably, the variable is an SX Phe pulsator blended 
with a red giant of about the same luminosity. The sinusoidal lightcurve of V157 has a stable period 
of $\sim$0.17 d and an amplitude of only $\sim0.015$ mag. If the marginally acceptable period two 
times longer were the proper one, then the star would be a blend of another red giant with a W UMa 
binary. V158 is a main-sequence object with $\Delta V=0.16$~mag and $P=0.2$ d. Again, if a period 
two times longer were the right one, another (unblended) W UMa would be the source of observed 
variations. Yet another W UMa, blended with a red giant/subgiant, is the only plausible explanation 
of regular variations with $P=0.39$ d observed in V159 - an object located at the base of the red
giant branch. The photometry of V160 may be affected by a much brighter star located next to this 
object. The fact that its slow brightening is clearly seen in light curves obtained both with DIAPL 
and DAOPHOT caused us to mark V160 as a suspected variable, although physical mechanism driving 
such a secular change in an otherwise normal turnoff star remains obscure.

VN11, VN12, VN15, VN16, VN18 and VN21 are likely spotted background or foreground objects (VN18 coincides 
with an X-ray source). The remaining stars VN13, VN14, VN16, VN17, VN19 and VN20 are located at the edge 
of our field of view, far away from the center of NGC 3201. VN13 with $P=0.38$ d is most likely a 
foreground W UMa seen at a low inclination. VN14 also has a W UMa-like lightcurve, however the 
period (0.18 d) is too short for this class of binaries. A period two times shorter matches the 
data equally well. If that is the case then we are observing a strong reflection effect, and VN14 becomes 
a truly interesting object. The low amplitude ($\Delta V=0.02$~mag) sinusoidal lightcurve of VN17 
has no straightforward interpretation. Since we cannot exclude that the observed changes are spurious,
we marked this object as a suspected variable. VN19 has a very stable period of 0.092 d, and it is
probably a foreground $\delta$Scuti pulsator. The noisy lightcurve of VN20 does not allow for an 
unambiguous interpretation. With a marginally acceptable period two times longer than that given
in Table \ref{tab:vardata}, it might be a foreground W UMa seen at a very low inclination.

\Section{Summary}
 \label{sec:sum}
We have conducted a ten-year long photometric survey of the globular cluster NGC 3201 in a search 
for variable stars. A total of 49 variables plus 8 suspected variables were discovered, and 
multiseasonal light curves were compiled for another 95 variables that had been known before. For 
all observed variables periods were obtained. Three new eclipsing binaries 
and eight pulsating stars of SX Phe-type were found in the blue-straggler region. Twenty-four SX Phe pulsators
are now known in NGC 3201, many of them multiperiodic. An asteroseismological study of the whole 
sample should allow a determination of BS masses and provide important constraints on theories explaining 
the origin of these objects. Several detached eclipsing binaries are interesting targets for follow-up
spectroscopic studies from which parameters of the components can be derived together with age and 
distance of the cluster. Particularly valuable among these are V119 - a blue straggler most probably 
undergoing mass transfer from a less massive but more evolutionarily advanced secondary to the primary,
and VN8 - a pair of subdwarfs on a wide three-day orbit seen nearly edge-on, which may bring important 
information about late evolutionary stages of cluster binaries. Another interesting object is V149 - 
a binary composed most likely of two subdwarfs on a very tight orbit. 

Four variable objects in our field of view were found to coincide with X-ray sources. Three of them 
are chromospherically active stars; the fourth one is a quasar with $z\approx 0.5$ which will serve 
as a reference point for the determination of the absolute proper motion of NGC 3201.

\Acknow
{WN, BM, WP and MR were partly supported by the grant DEC-2012/05/B/ST9/03931
from the Polish National Science Center. We thank Grzegorz Pojma\'nski for the lc code which 
vastly facilitated the work with light curves. 
}


\section*{Appendix A: Period-changing W UMa stars}

\begin{figure}
   \centerline{\includegraphics[width=0.95\textwidth,
               bb = 46 320 566 688, clip]{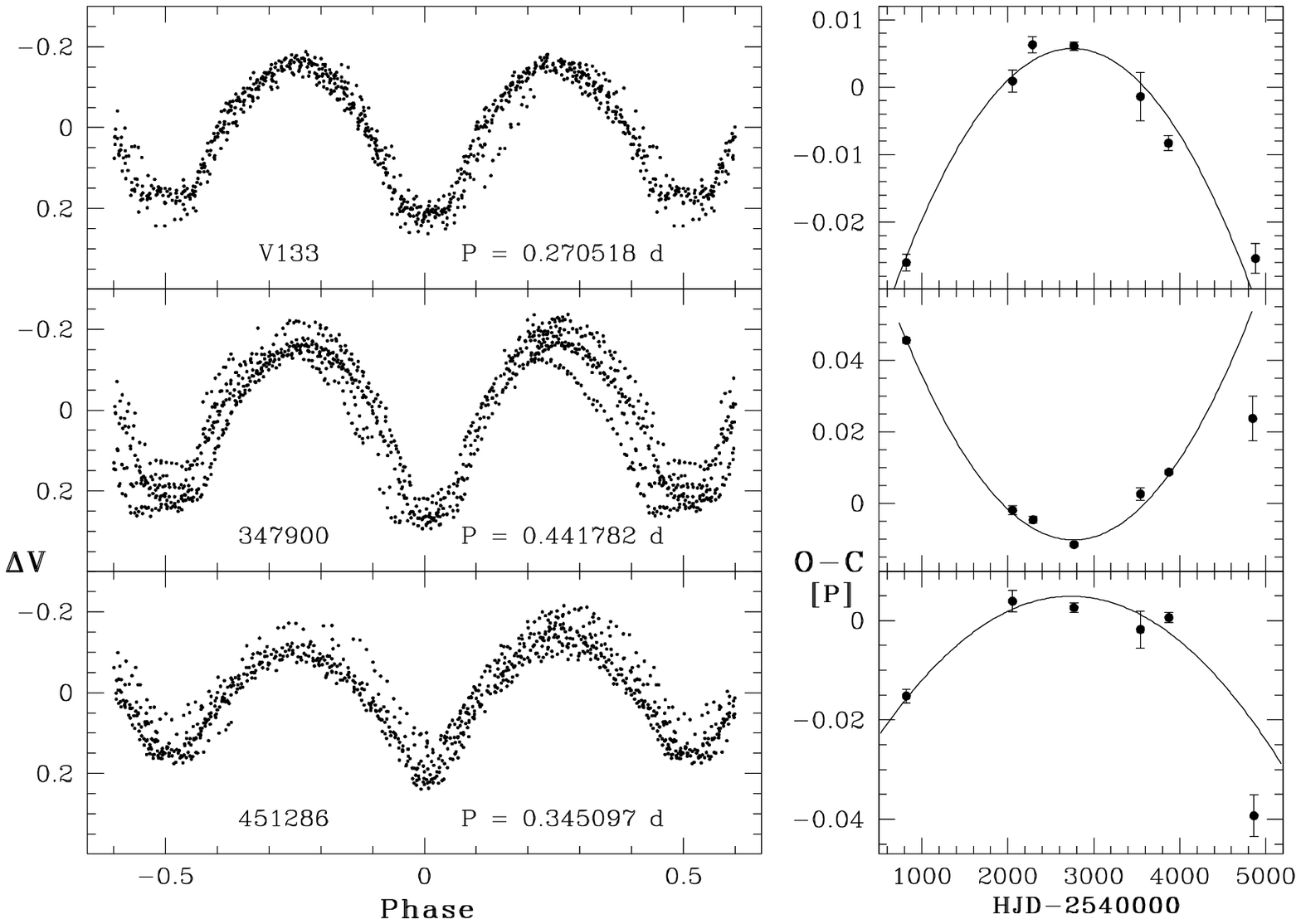}}
   \caption{Left column: light curves of period-changing W UMa binaries phased 
            with the best-fitting constant periods. Stars \#347900
            and \#451286 were discovered by von Braun \& Mateo (2002) who 
            listed them as V4 and V2, respectively. Right column: O-C diagrams
            for these systems. 
    \label{fig:wumas}}
\end{figure}

Attempts to fit the best period for the contact binary V133 revealed poor matching of light 
curves from different seasons. The same was found for another two such systems: \#347900
and \#451286 on our list of light curves, which had been discovered by von Braun \& Mateo (2002; 
their stars V4 and V2, respectively). Their light curves phased with the best-fitting periods
are shown in the left column of Fig. \ref{fig:wumas}. Investigations reported below were 
conducted to check for possible period changes. 

For each star first we determined the best orbital frequency $\nu$ from the multiharmonic 
periodogram (Schwarzenberg-Czerny 1996, 2012). In the process, by means of orthogonal projection
all observations were fitted with a series of Szeg\"o (1939) trigonometric orthogonal 
polynomials $p_n(z)$ of order $2N=16$:
\begin{equation}
f(t)=z^{-N}\sum_{n=0}^{2N}c_np_n(z)\label{ea.1}
\end{equation} 
where $z=e^{2\pi i\nu t}$. This real series served as our template lightcurve of fixed shape 
and amplitude. Next we took sets of observations from each season and by nonlinear least 
squares (NLSQ) we fitted them with our template by adjusting phase and zero point. As formulae
for differentiation of Szeg\"o polynomials are not readily available, we list them 
in Appendix B.

To protect the NLSQ solution from effects of poor phase coverage we inspected the correlation 
coefficient $\rho$ between phase and zero point. Whenever $|\rho|$ exceeded $0.3$, the 
two-parameter fit was replaced with a single-parameter one (the phase shift was only fitted). 
This typically occured for the two worst-populated seasons. The zero shifts did not exceed 
0.06 mag and usually were a fraction of that. Their cause is not clear; both blending and 
intrinsic variations are plausible explanations. Throughout the calculations we accounted 
for weights derived from the errors of observations. 
The typical scatter of observations was 0.02 mag.

The results of fitting are plotted in the right column of Fig. \ref{fig:wumas} in the form of 
O-C diagrams. An inspection of plots reveals large systematic phase shifts. To further 
investigate the trend we fitted seasonal points with parabolae. For V133 and \#451286 fits 
were satisfactory, judging from $\chi^2$ of order 1 per degree of freedom. For \#347900 
$\chi^2(4)= 6.9$ suggests possibly variable rate of period change. The quadratic ephemerides 
of primary eclipses 
\begin{equation}
HJD(Min)=  T_0 +  P\;E + \frac{1}{2}\dot{P}\;E^2                                                         
\end{equation} 
corresponding to these parabolas are listed in Table \ref{ta.1}.
\begin{table}[H]
\caption{Moments of primary minima}\label{ta.1}
\begin{center}
\begin{tabular}{|r|r|c|c|}
\hline
Name & T$_0$ & P & $\frac{1}{2}\dot{P}$\\ \hline
V133 & 2763.69610 & 0.270517684 &  1.6e-10\\
$\pm$ & 0.00032 & 0.000000023  &  1.5e-11\\
347900 & 2765.64121 & 0.441782086 & -1.3e-09\\
$\pm$ &   0.00057 & 0.000000108 &   8.0e-11\\
451286 & 2766.70008 &  0.345096870 &  2.3e-10\\
 $\pm$ &   0.00101 & 0.000000084   & 6.7e-11\\ \hline
\end{tabular} 
\end{center}
\end{table}
Our results reveal period changes of both signs, on time scales $P/\dot{P}$ ranging
from $-5\times10^5$ to $2\times10^6$ y, consistent with thermal time scales of binary components.
Since the early ideas of Lucy (1968a, 1968b) the contact binaries are known to suffer from thermal 
instability due to their marginal contact. As these ideas were further elaborated by 
Flannery(1976), Lucy(1976) and Robertson \& Eggleton(1977), a picture emerged of thermal 
relaxation oscillations (TRO), leading to alternate breaking and 
re-establishing of the contact. At some phases energy and mass could flow
between the components. In particular, the period of V133 is increasing on a time scale of 
$2.3\times10^6$y, indicating an active contact phase. While there is no evidence of orbital momentum 
change at this stage, stellar expansion or contraction can still lead to the corresponding period 
changes. Evolutionary scenarios invoke angular momentum loss during formation of W UMa systems, either 
during common envelope phase or due to magnetic wind (e.g. Stepien \& Gazeas, 2012).

\section*{Appendix B: Differentiation of Szeg\"o orthogonal polynomials}

Expansion of orthogonal polynomials into monomials constitutes an ill posed
problem, Vandermonde matrices beeing one example. Hence instead of trigonometric functions
we employ direct derivatives of Szeg\"o recurrence relation Eq. (\ref{e2.11}), 
in the process avoiding $nz^{n-1}$ for numerical stability.
Calculations can be performed by means of the recurrence starting from:
\begin{eqnarray}
p_0(z)&=&1\label{e2.6}\\
p_0(z)'&=&0\label{e2.7}\\
p_0(z){\dagger}'&=&0\label{e2.8}
\end{eqnarray}
and continuing for $n=0,1,\cdots,2N$
\begin{eqnarray}
c_n&=&\frac{(p_n,z^Nx)}{\|p_n\|^2}\label{e2.9}\\
\alpha_n&=&\frac{(1,zp_n)}{\|p_n\|^2}\label{e2.10}\\
p_{n+1}&=&zp_n-\alpha_np_n{\dagger}\label{e2.11}\\
p_{n+1}{\dagger}&=&p_n{\dagger}-\overline{\alpha_n}zp_n\label{e2.12}\\
w&=&z(ip_n+p_n')\label{e2.13}\\
p_{n+1}'&=&w-\alpha_np_n{\dagger}'\label{e2.14}\\
p_{n+1}{\dagger}'&=&p_n{\dagger}'-\overline{\alpha_n}w\label{e2.15}
\end{eqnarray}
where for $|z|=1$ dagger denotes reversion of polynomial coefficients 
$p_n(z)\dagger=z^n\overline{p_n(z)}$ and
apostrophe denotes derivative over phase $\varphi$, so that $z'=(e^{i\varphi})'=iz$.
The scalar product is derived from times $t_m$ and weights $w_m\geq 0$, $m=1,\cdots,M$
of observations $x_m$: $(f,g)=\left.\sum_{m=1}^Mw_m\overline{f(z_m)}g(z_m)\right|_{z_m=e^{2\pi i\nu t_m}}$.

\clearpage
\begin{center}
\footnotesize{
\begin{longtable}{|c|c|c|c|c|c|c|l|c|}
\caption[]{Basic data of NGC 3201 variables discovered
within the present survey$^*$}\label{tab:vardata}\\
  \hline
 Id & RA & DEC & $V$ & $B-V$ & $\Delta V$& Period  & Type$^a$& Mem$^b$ \\
    & [deg] & [deg]  &[mag]& [mag] & [mag]    & [d]     & Remarks & \\
  \hline
\endfirsthead
\multicolumn{9}{c}
{{\tablename\ \thetable{} concluded}} \\
  \hline
 Id & RA & DEC & $V$ & $B-V$ & $\Delta V$& Period  & Type& Mem$^b$ \\
    & [deg] & [deg]  &[mag]& [mag] & [mag]    & [d]     &Remarks   & \\
  \hline
\endhead
\hline
\multicolumn{9}{p{12.1 cm}}
{$^*$We follow the naming convention of Clement et al. (2001; 2012 edition) continued by Arellano 
Ferro et al. (2014), whose last cataloged star was V124. V119 is additionally listed as a system 
for which we provide important new information. Stars VN1 -- VN22 are either not PM-members of NGC 
2301 or PM data are missing for them.}\\
\multicolumn{9}{p{12.1 cm}}
{$^a$EA: detached eclipsing binary, EB: close eclipsing binary, EW: contact eclipsing binary, 
SX:~SX~Phe-type pulsator, SP: spotted variable, LPV: long-period variable, ?: roughly sinusoidal
variations of unknown nature, BS: blue straggler, RG: red giant.}\\ 
\multicolumn{9}{p{12.1 cm}}
{$^b$Y: member, N: nonmember, U: no data or data ambiguous.}\\
\multicolumn{9}{p{12.1 cm}}
{$^1$Period taken from the unpublished PhD thesis of BM. $^2$Periodicity may be spurious. 
$^3$Irregular variations are also possible. $^4$Clear periodicity observed in May 2003 only. 
$^5$Period is the time-span between the first and the last observation.}\\
\endfoot
\hline
\endlastfoot
V119 & 154.43733 & -46.42035 & 16.049 & 0.582 & 0.202 & 2.551572 & EA,BS & U\\
V125 & 154.53096 & -46.36382 & 17.449 & 0.512 & 0.045 & 0.042236 & SX,BS & Y\\
V126 & 154.43381 & -46.34955 & 17.405 & 0.473 & 0.052 & 0.039888 & SX,BS & Y\\
V127 & 154.41949 & -46.40930 & 17.184 & 0.440 & 0.064 & 0.045335 & SX,BS & Y\\
V128 & 154.40258 & -46.39864 & 17.410 & 0.495 & 0.052 & 0.034489 & SX,BS & Y\\
V129 & 154.39888 & -46.38365 & 17.172 &  --   & 0.051 & 0.051988 & SX & Y\\
V130 & 154.41882 & -46.42159 & 16.357 & 0.477 & 0.021 & 0.047317 & SX,BS & Y\\
V131 & 154.31400 & -46.38674 & 17.244 & 0.462 & 0.041 & 0.033090 & SX,BS & Y\\
V132 & 154.26829 & -46.33185 & 16.551 & 0.430 & 0.103 & 0.047643 & SX,BS & Y\\
V133 & 154.43623 & -46.38333 & 18.305 & 0.645 & 0.369 & 0.270517 & EW & Y\\
V134 & 154.43366 & -46.40117 & 17.520 & 0.519 & 0.094 & 0.285316 & EW,BS & Y\\
V135 & 154.41299 & -46.40830 & 17.916 & 0.788 & 0.130 & 0.279262 & EW & Y\\
V136 & 154.35266 & -46.41218 & 16.197 & 0.352 & 0.033 & 0.702218 & EW,BS & Y\\
V137 & 154.42977 & -46.37028 & 17.911 & 0.653 & 0.087 & 7.429154$^1$ & EA & Y\\
V138 & 154.36805 & -46.39069 & 18.292 & 0.720 & 0.124 & 9.199380 & EA & Y\\
V139 & 154.34408 & -46.39947 & 18.076 & 0.716 & 0.297 & 8.789090 & EA & Y\\
V140 & 154.31382 & -46.41273 & 18.499 & 0.801 & 0.411 &     --   & EA & Y\\
V141 & 154.38809 & -46.41872 & 17.154 & 0.779 & 0.149 & 10.00370 & EA & Y\\
V142 & 154.29286 & -46.37735 & 17.257 & 0.756 & 0.108 & 27.69300 & EA & Y\\
V143 & 154.28997 & -46.46769 & 20.565 & 1.049 & 0.324 & 0.380726 & EA & Y\\
V144 & 154.55648 & -46.32846 & 14.626 & 1.066 & 0.025 & 0.938225$^2$ & EB? & Y\\
V145 & 154.45097 & -46.48473 & 18.133 & 0.679 & 0.110 & 24.34624 & ? & Y\\
V146 & 154.37700 & -46.41179 & 17.275 & 0.932 & 0.158 & 13.75904$^1$ & SP,RG & Y\\
V147 & 154.33856 & -46.43918 & 17.410 & 0.757 & 0.124 & 1.910503$^3$ & SP? & Y\\
V148 & 154.54247 & -46.31482 & 17.235 & 0.889 & 0.040 & 0.467678$^4$ & ? & Y\\
V149 & 154.50747 & -46.36550 & 21.280 & 0.701 & 0.485 & 0.062268 & EB? & Y\\
V150 & 154.51355 & -46.39422 & 15.225 & 0.322 & 0.016 & 0.18642$^2$ & ? & Y\\
V151 & 154.40944 & -46.33032 & 14.275 & 1.043 & 0.021 & 1.033544$^3$ & ? & Y\\
V152 & 154.39568 & -46.32713 & 14.692 & 0.756 & 0.018 & 0.516789 & ? & Y\\
V153 & 154.41524 & -46.36521 & 17.592 & 0.484 & 0.089 & 0.899928$^3$ & ? & Y\\
V154 & 154.36395 & -46.39678 & 16.498 & 0.321 & 0.024 & 0.168045$^4$ & ? & Y\\
V155 & 154.35480 & -46.42202 & 17.690 & 0.670 & 0.056 & 0.252956 & EW? & Y\\
V156 & 154.39468 & -46.44167 & 17.084 & 0.840 & 0.029 & 0.041316 & SX? & Y\\
V157 & 154.38582 & -46.44596 & 15.651 & 0.916 & 0.016 & 0.174603 & EW? & Y\\
V158 & 154.34077 & -46.44567 & 19.501 & 0.805 & 0.162 & 0.199170 & EW? & Y\\
V159 & 154.29128 & -46.38593 & 17.392 & 0.894 & 0.028 & 0.387159 & EW? & Y\\
V160 & 154.44577 & -46.48059 & 17.695 & 0.611 & 0.170 & 2100.000$^5$ & LPV & Y\\
VN1 & 154.41641 & -46.39019 & 17.682 & 0.856 & 0.113 & 0.398926 & EW & N\\
VN2 & 154.40006 & -46.42012 & 17.702 & 0.533 & 0.359 & 0.297544 & EW,BS? & N\\
VN3 & 154.43425 & -46.47440 & 19.660 & 0.924 & 0.384 & 0.336212 & EW & U\\
VN4 & 154.40413 & -46.44006 & 21.713 &  --   & 0.905 & 0.398336 & EW & U\\
VN5 & 154.27474 & -46.34356 & 17.197 & 0.785 & 0.154 & 0.286946 & EW & N\\
VN6 & 154.54876 & -46.31227 & 18.576 & 0.868 & 0.104 & 1.063468 & EA & N\\
VN7 & 154.53385 & -46.45677 & 20.570 & 1.430 & 0.464 & 0.967542 & EA & N\\
VN8 & 154.42951 & -46.37056 & 20.641 & 0.602 & 0.693 & 2.860388 & EA & U\\
VN9 & 154.45077 & -46.44544 & 16.441 & 0.759 & 0.087 & 8.909326 & EA & N\\
VN10 & 154.27129 & -46.56679 & 17.641 & 0.924 & 0.326 &    --    & EA & N\\
VN11 & 154.43976 & -46.45586 & 17.521 & 0.581 & 0.111 & 4.341275 & SP & N\\
VN12 & 154.49269 & -46.47945 & 14.355 & 0.922 & 0.026 & 23.84405 & SP & N\\
VN13 & 154.50828 & -46.26136 & 15.780 & 0.846 & 0.019 & 0.377768 & EW? & N\\
VN14 & 154.49405 & -46.21687 & 19.967 & 1.502 & 0.389 & 0.088603 & EW? & U\\
VN15 & 154.46060 & -46.28940 & 17.623 & 1.255 & 0.140 & 15.95484 & SP? & N\\
VN16 & 154.56195 & -46.36209 & 16.075 & 0.884 & 0.030 & 1.123061 & SP? & N\\
VN17 & 154.55929 & -46.38112 & 15.375 & 0.886 & 0.022 & 0.493302 & ? & N\\
VN18 & 154.52895 & -46.34859 & 16.566 & 1.233 & 0.035 & 7.862955 & SP? & N\\
VN19 & 154.55825 & -46.45848 & 14.668 & 0.635 & 0.030 & 0.092057 & ? & N\\
VN20 & 154.59523 & -46.48295 & 14.653 & 0.596 & 0.016 & 0.166173 & ? & U\\
VN21 & 154.30577 & -46.37241 & 18.590 & 1.179 & 0.084 & 7.299531 & SP? & N\\
 \end{longtable}
}
\end{center}


\begin{subfigures}
\begin{figure}
   \centerline{\includegraphics[width=0.95\textwidth,
               bb = 287 175 992 737, clip]{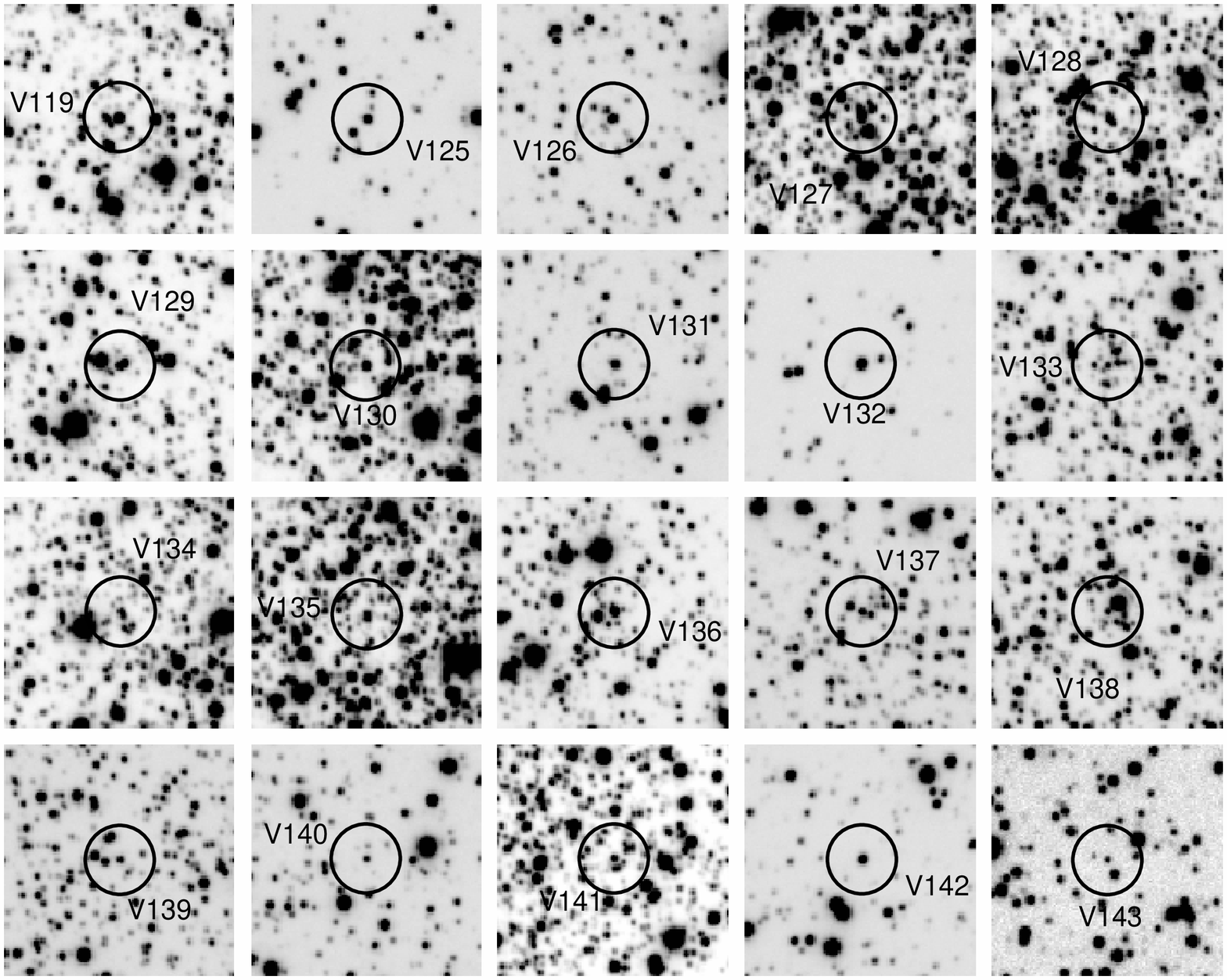}}
   \vspace{1.5 mm}
   \centerline{\includegraphics[width=0.95\textwidth,
               bb = 287 175 992 737, clip]{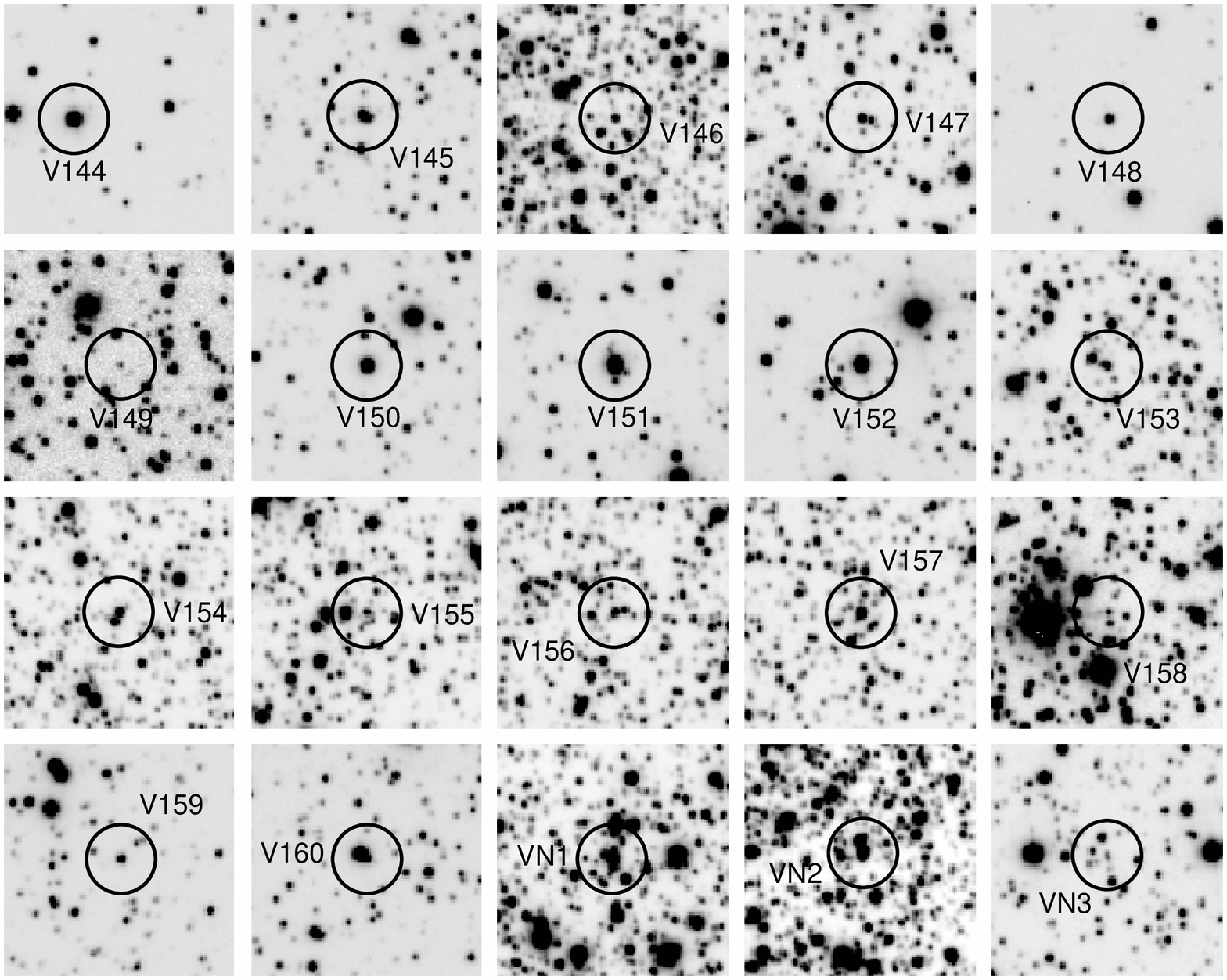}}
\caption{Finding charts for the variables.
    Each chart is 30 arcsec on a side; north is up and east to the left.
    \label{fig:maps1}}
\end{figure}

\begin{figure}
   \centerline{\includegraphics[width=0.95\textwidth,
               bb = 287 175 992 737, clip]{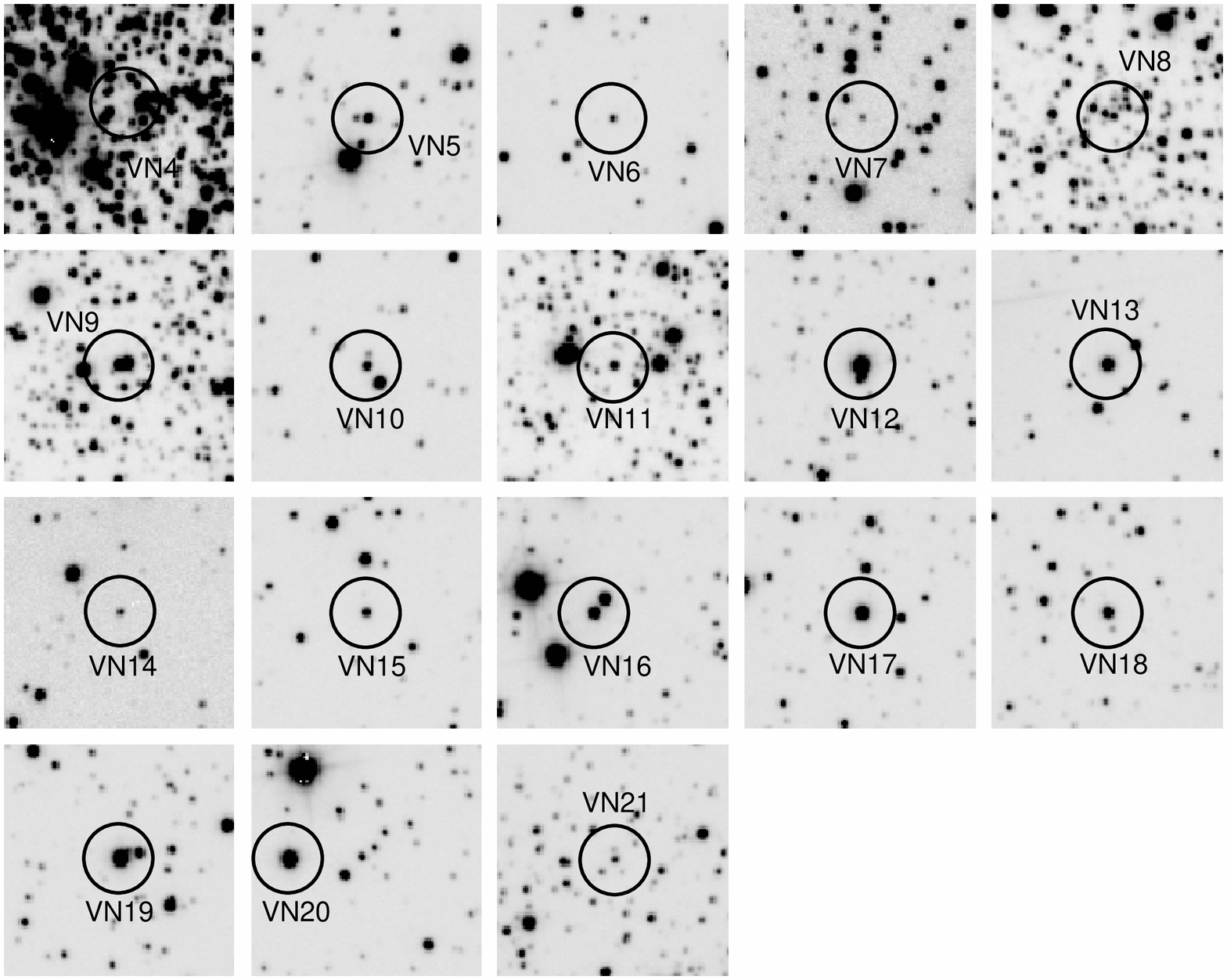}}
\caption{Continuation of Fig. \ref{fig:maps1}.
    \label{fig:maps2}}
\end{figure}
\end{subfigures}

\begin{subfigures}
\begin{figure}
   \centerline{\includegraphics[width=0.95\textwidth,
               bb = 42 26 528 767, clip]{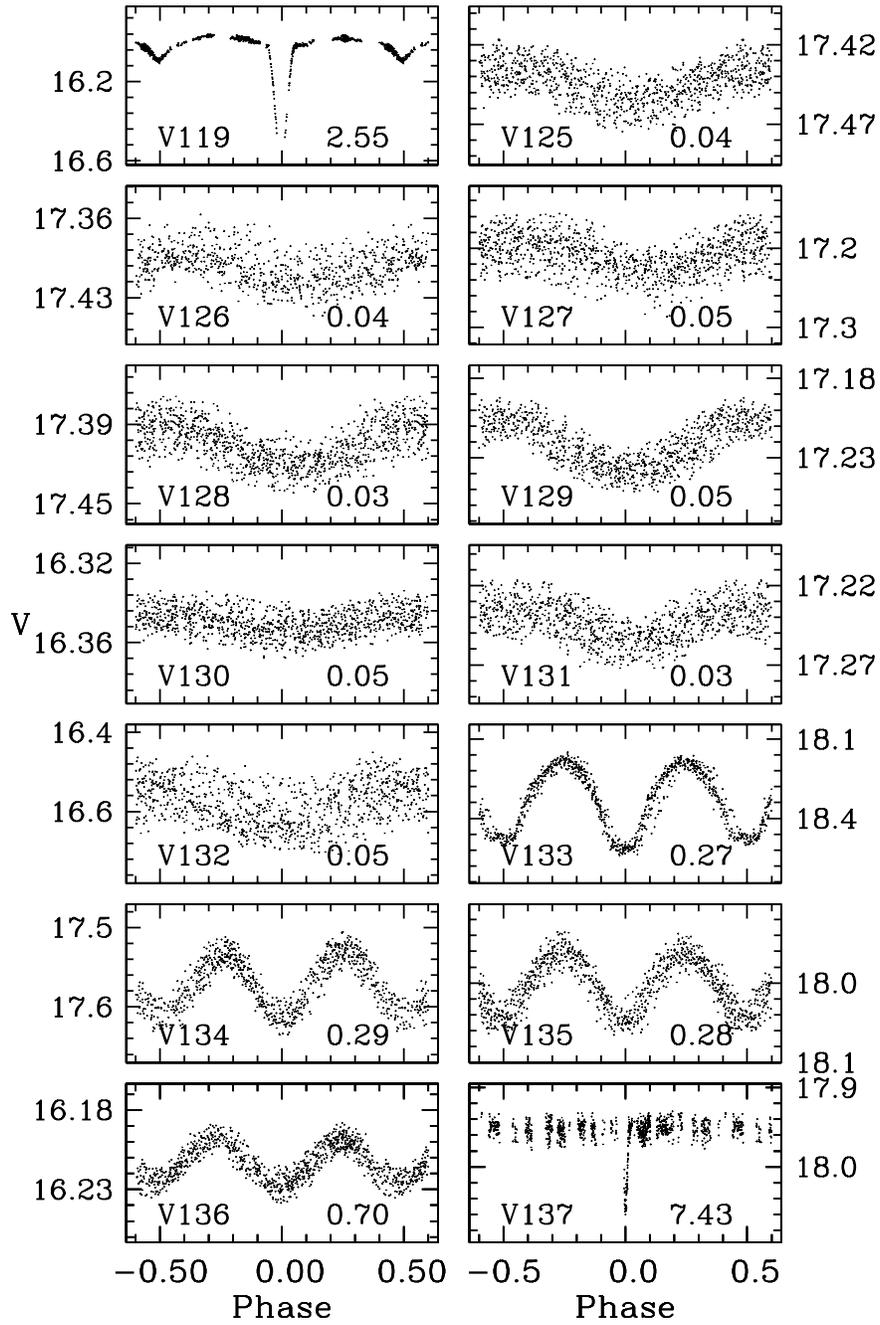}}
   \caption{Phased $V$ curves of variables detected in the field of
    NGC3201. Inserted labels give star ID and period in days.
    \label{fig:curves1}}
\end{figure}

\begin{figure}
   \centerline{\includegraphics[width=0.95\textwidth,
       bb = 42 26 528 765, clip]{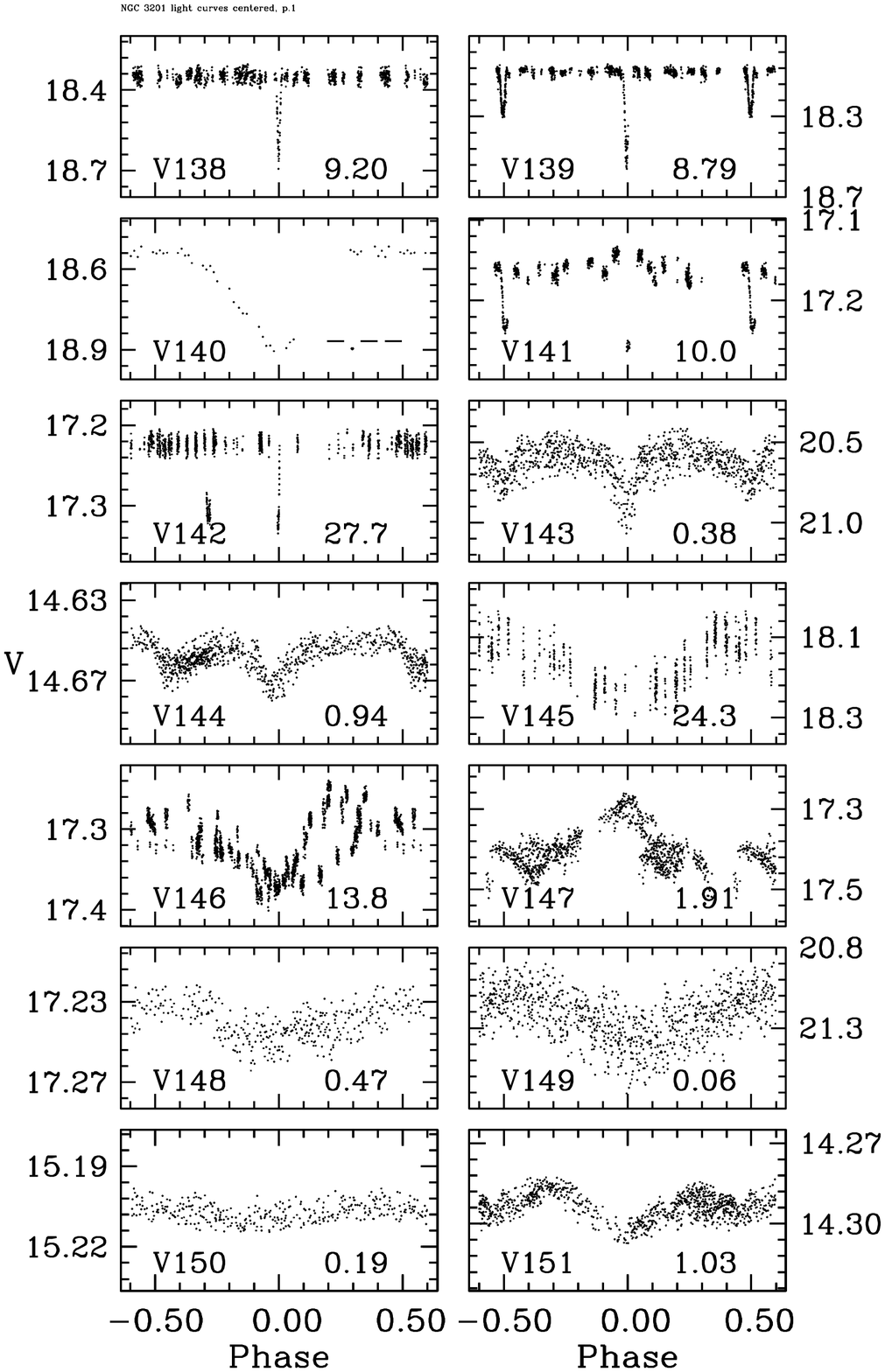}}
   \caption{Continuation of Fig. \ref{fig:curves1}.  
    \label{fig:curves2}}
\end{figure}

\begin{figure}
   \centerline{\includegraphics[width=0.95\textwidth,
       bb = 42 26 528 765, clip]{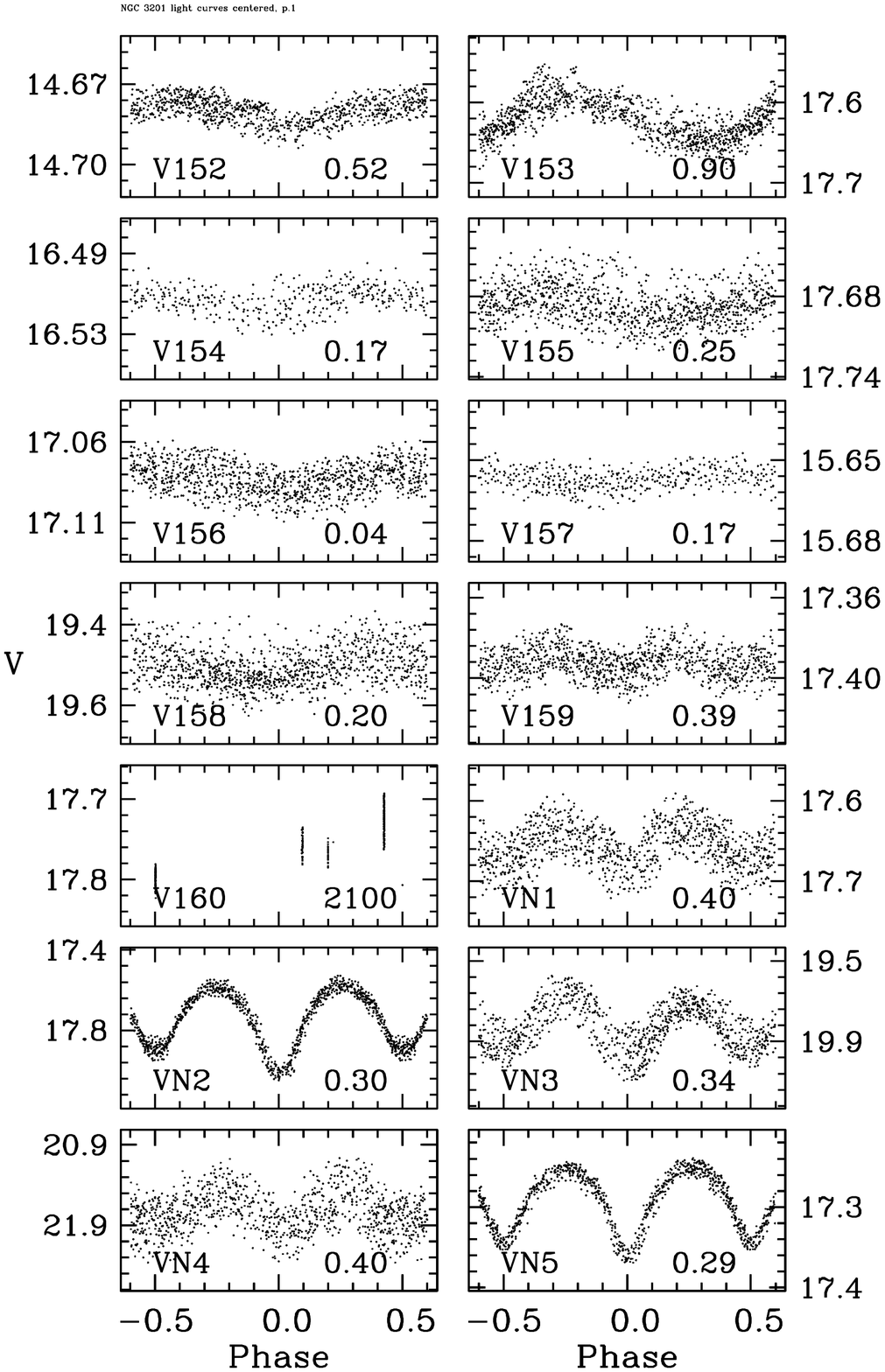}}
   \caption{Continuation of Fig. \ref{fig:curves1}.  
    \label{fig:curves3}}
\end{figure}

\begin{figure}
   \centerline{\includegraphics[width=0.95\textwidth,
       bb = 42 26 528 765, clip]{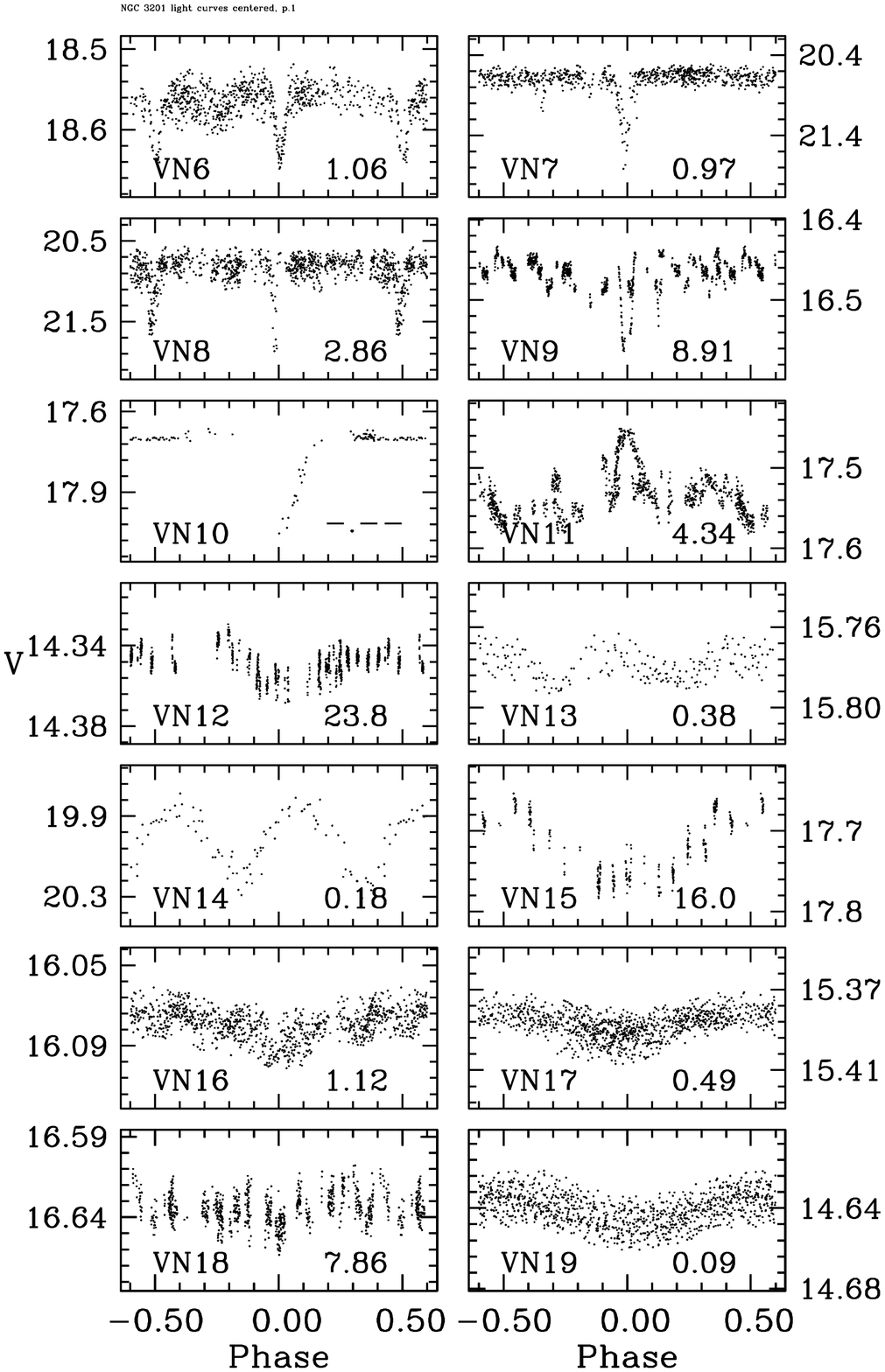}}
   \caption{Continuation of Fig. \ref{fig:curves1}.  
    \label{fig:curves4}}
\end{figure}

\begin{figure}
   \centerline{\includegraphics[width=0.95\textwidth,
       bb = 42 631 528 765, clip]{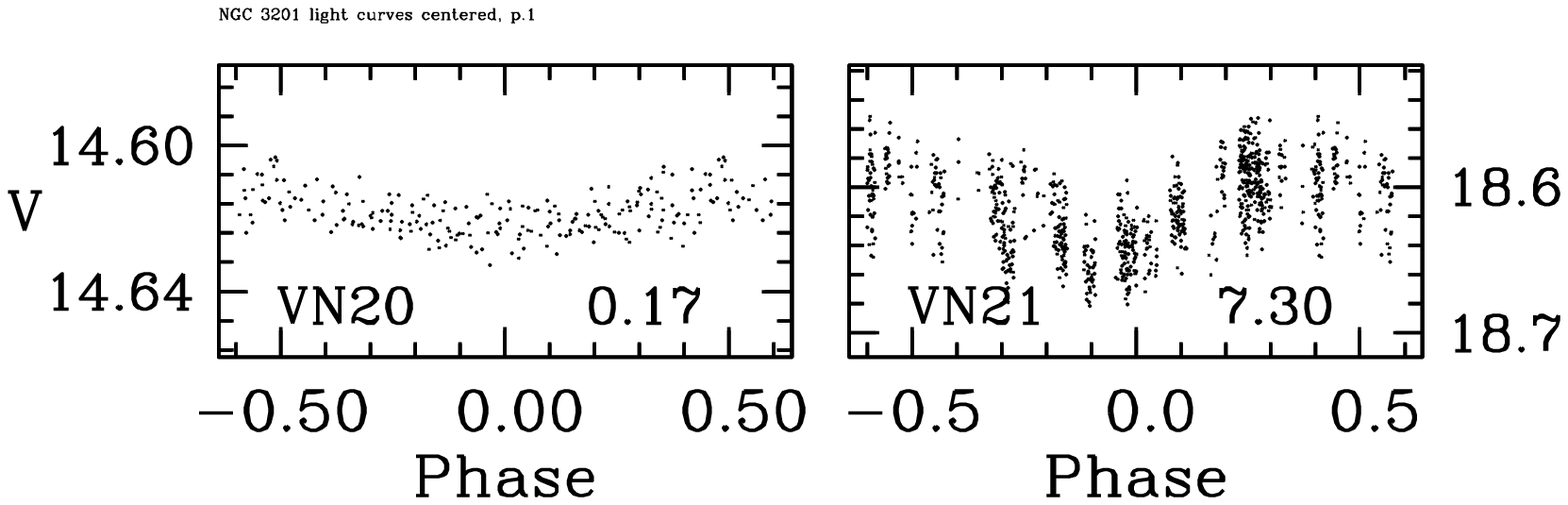}}
   \caption{Continuation of Fig. \ref{fig:curves1}.  
    \label{fig:curves5}}
\end{figure}
\end{subfigures}

\end{document}